\documentclass[12pt,english,a4paper]{article}
\usepackage[T1]{fontenc}
\usepackage[latin9]{inputenc}
\usepackage{amsmath}
\usepackage{graphicx}

\makeatletter

\usepackage{float}

\usepackage{epsfig}

\usepackage{babel}
\makeatother

\begin{document}

\makeatother

\title{\textbf{\large Symmetry preserving regularization with a cutoff}}

\author{G. Cynolter and E. Lendvai}

\date{Theoretical Physics Research Group of Hungarian Academy of Sciences,
Eötvös University, Budapest, 1117 Pázmány Péter sétány 1/A, Hungary }

\maketitle
\begin{abstract}
A Lorentz and gauge symmetry preserving regularization method is proposed
in 4 dimension based on momentum cutoff. We use the conditions of
gauge invariance or freedom of shift of the loop-momentum to define
the evaluation of the terms carrying Lorentz indices, e.g. proportional
to $k_{\mu}k_{\nu}$. The remaining scalar integrals are calculated
with a four dimensional momentum cutoff. The finite terms (independent
of the cutoff) are unambiguous and agree with the result of dimensional
regularization.
\end{abstract}

\section{Introduction}

Several regularization methods are known and used in quantum field
theory: three and four dimensional momentum cutoff, Pauli-Villars
type, dimensional regularization, lattice regularization, Schwinger's
proper time method and others directly linked to renormalization like
differential renormalization. Dimensional regularization (DREG) \cite{dreg}
is the most popular and most appreciated as it respects the gauge
and Lorentz symmetries of the Lagrangian and textbooks give a detailed
recipe. However DREG is not useful in all cases, for example it is
not directly applicable to supersymmetric gauge theories as it modifies
the number of bosons and fermions differently. DREG gets rid of (does
not identify) naive quadratic divergencies, which may be important
in low energy effective theories or in the Wilson's renormalization
group method. Another shortcoming is that together with (modified)
minimal subtraction DREG is a {}``mass independent'' scheme, particle
thresholds and decoupling must put in the theory by hand \cite{georgi}.
The choice of the ultraviolet regulator always depends on the problem.

In low energy effective field theories or in the Wilson renormalization
group method there is an explicit cutoff, with well defined physical
meaning. The cutoff gives the range of validity of the model. There
are a few implementations: sharp momentum cutoff in 3 or 4 dimensions,
modified operator regularization (based on Schwinger proper time method
\cite{schwinger}). In the Nambu-Jona-Lasinio model different regularizations
proved to be useful calculating different physical quantities \cite{klev}. 

Regularization is an arbitrary algorithm that defines how to handle
divergent momentum integrals. In this paper we show that with a reasonable
and definite modification the loop calculations can be reduced to
scalar integrals and those can be evaluated with a sharp momentum
cutoff. The results respect gauge (chiral and other) symmetries. Using
a naive momentum cutoff the symmetries are badly violated. The calculation
of the QED vacuum polarization function ($\Pi_{\mu\nu}(q)$) shows
the problems. The Ward identity tells us that $q^{\mu}\Pi_{\mu\nu}(q)=0$,
e.g. in\begin{equation}
\Pi_{\mu\nu}(q)=q_{\mu}q_{\nu}\Pi_{L}(q^{2})-g_{\mu\nu}q^{2}\Pi_{T}(q^{2})\label{eq: Pidef}\end{equation}
 the two coefficients must be the same $\Pi(q^{2})$. Usually the
condition $\Pi(0)=0$ is required to define a subtraction to keep
the photon massless at 1-loop. However this condition is ambiguous
when one calculates at $q^{2}\neq0$ in QED or in more general models,
in the case of two different masses in the loop, it just fixes $\Pi(q^{2},\ m_{1},\ m_{2})$
in the limit of degenerate masses at $q^{2}=0$. Ad hoc subtractions
does not necessarily give satisfactory results.

There were several proposals to define symmetry preserving cutoff
regularization. Usual way is to start with a regularization that respects
symmetries and find the connection with momentum cutoff. In case of
dimensional regularization already Veltman observed \cite{veltman}
that the naive quadratic divergencies can be identified with the poles
in two dimensions (d=2) besides the usual logarithmic singularities
in d=4. This idea turned out to be fruitful. Hagiwara et al. \cite{hagiwara}
calculated electroweak radiative corrections originating from effective
dimension-six operators and later Harada and Yamawaki performed the
Wilsonian renormalization group inspired matching of effective hadronic
field theories \cite{harada}. Based on Schwinger's proper time approach
Oleszczuk proposed the operator regularization method \cite{Olesz},
and showed that it can be formulated as a smooth momentum cutoff respecting
gauge symmetries \cite{Olesz,liao}. A momentum cutoff is defined
in the proper time approach in \cite{varin} with the identification
under loop integrals \begin{equation}
k_{\mu}k_{\nu}\rightarrow\frac{1}{d}g_{\mu\nu}k^{2}\label{eq: perdim}\end{equation}
instead %
\footnote{In what follows we denote the metric tensor by $g_{\mu\nu}$ both
in Minkowski and Euclidean space.%
} of the standard $d=4$. The degree of the divergence determines $d$
in the result: $\Lambda^{2}$ goes with $d=2$ and $\ln(\Lambda^{2})$
with $d=4$. This way the authors get correctly the\textit{ divergent
parts,} they checked them in the QED vacuum polarization function
and in the phenomenological chiral model.

Various authors formulated consistency conditions to maintain gauge
invariance during the evaluation of divergent loop integrals. Finite
\cite{gu} or infinite \cite{horejsi,wu1} number of new regulator
terms added to the propagators a'la Pauli-Villars, the integrals are
tamed to have at most logarithmic singularities and become tractable.
Differential renormalization can be modified to fulfill consistency
conditions automatically, it is called constrained differential renormalization
\cite{cdr}. Another method, later proved to be equivalent with the
previous one \cite{equiv}, is called implicit regularization, a recursive
identity (similar to Taylor expansion) is applied and the external
momentum ($q$) is moved to finite integrals. The divergent integrals
contain only the loop momentum, thus universal local counter terms
can cancel the potentially dangerous symmetry violating contributions
\cite{nemes,nemes2}. Gauge invariant regularization is implemented
in exact renormalization group method providing a cutoff without gauge
fixing in \cite{rosten}. Introducing a multiplicative regulator in
the d-dimensional integral, the integrals are calculable in the original
dimension with the tools of DREG \cite{dillig}.

We show that there is a tension between naive application of the Lorentz
symmetry and gauge invariance. The proper handling of the $k_{\mu}k_{\nu}$
terms in divergent loop-integrals solve the problems of momentum cutoff
regularizations. We give a simple and well defined algorithm to have
unambiguous finite and infinite terms, the finite terms agree with
the result of DREG. 

In section 2 we present how to get a momentum cutoff from DREG calculation,
then we give the gauge symmetry preserving conditions emerging during
the calculation of the vacuum polarization amplitude. In section 4
we discuss the condition of independence of momentum routing in loop
diagrams. Section 5 shows that gauge invariance and freedom of shift
in the loop momentum have the same root. Next we show that the conditions
are related to vanishing integrated surface terms. In section 7 we
give a definition of the new regularization method and in section
8 as an example we present the calculation of a general vacuum polarization
function at 1-loop and close with conclusions.

\section{Momentum cutoff via dimensional regularization}

DREG is very efficient and popular, because it preserves gauge and
Lorentz symmetries. Performing standard steps the integrals evaluated
in $d=4-2\epsilon$ dimension. Generally the loop-momentum integral
Wick rotated and with a Feynman parameter ($x$) the denominators
are combined, then the order of $x$ and momentum integrals are changed.
Shifting the loop-momentum does not generate surface terms and it
leads to spherically symmetric denominator, terms linear in the momentum
are dropped and \eqref{eq: perdim} is used. Singularities identified
as $1/\epsilon$ poles, naive power counting shows that these are
the logarithmic divergencies of the theory.%
\footnote{Similar identification can be done in three dimensional integrals,
too \cite{anselmi}.%
} In DREG quadratic or higher divergencies are set identically to zero.
However Veltman noticed \cite{veltman} that quadratic divergencies
can be calculated in $d=2-2(\epsilon-1)$ in the limit $\epsilon\rightarrow1$.
This observation led to a cutoff regularization based on DREG.

Carefully calculating the one and two point Passarino-Veltman functions
in DREG and in 4-momentum cutoff the divergencies can be matched as
\cite{hagiwara,harada} \begin{eqnarray}
4\pi\mu^{2}\left(\frac{1}{\epsilon-1}+1\right) & = & \Lambda^{2},\label{eq: quad}\\
\frac{1}{\epsilon}-\gamma_{E}+\ln\left(4\pi\mu^{2}\right)+1 & = & \ln\Lambda^{2},\label{eq:log}\end{eqnarray}
where $\mu$ is the mass-scale of dimensional regularization. The
finite part of a divergent quantity is defined as \begin{equation}
f_{{\rm finite}}=\lim_{\epsilon\rightarrow0}\left[f(\epsilon)-R(0)\left(\frac{1}{\epsilon}-\gamma_{E}+\ln4\pi+1\right)-R(1)\left(\frac{1}{\epsilon-1}+1\right)\right],\label{eq:finite}\end{equation}
where $R(0)$, \, $R(1)$ are the residues of the poles at $\epsilon=0,\,1$
respectively. Note that in the usual $\epsilon\rightarrow0$ limit
the left hand side (lhs) of \eqref{eq: quad} vanishes and no quadratic
divergence appears in the original DREG.

The identifications above define a momentum cutoff calculation based
on the symmetry preserving DREG formulae. This cutoff regularization
is well defined and unique, but still relies on DREG. Let us see the
main properties in the calculation of the vacuum polarization function.
In $\Pi_{\mu\nu}$ the quadratic divergence is partly coming from
a $k_{\mu}k_{\nu}$ term via $\frac{1}{d}\cdot g_{\mu\nu}k^{2}$,
which is evaluated at $d=2$ instead of the $d=4$ in the naive cutoff
calculation. The $\Lambda^{2}$ terms cancel if and only if this term
is evaluated at $d=2$. This is a warning that the usual $k_{\mu}k_{\nu}\rightarrow\frac{1}{4}g_{\mu\nu}k^{2}$
substitution during the naive cutoff calculation of divergent integrals
might be too naive, especially as an intermediate step, the Wick rotation
is legal only for finite integrals. A further finite term additional
to the logarithmic singularity is coming from the well known expansion
in $\frac{1}{4-2\epsilon}\frac{1}{\epsilon}\simeq\frac{1}{4}\left(\frac{1}{\epsilon}+\frac{1}{2}\right)$,
and it is essential to retain gauge invariance. We stress that the
shift of the loop momentum is allowed in DREG, an improved cutoff
regularization should inherit it. In the next sections we derive consistency
conditions for general regularizations.

\section{Consistency conditions - gauge invariance}

Calculation in a gauge theory ought to preserve gauge symmetries.
We start with a single example, calculate the QED vacuum polarization
function with massive electrons, and present the condition(s) of gauge
invariance. We start generally (see Fig. 1.) with two different masses
\cite{fcmlambda} and restrict it to QED later.

\begin{figure}
\begin{centering}
\includegraphics[scale=0.5]{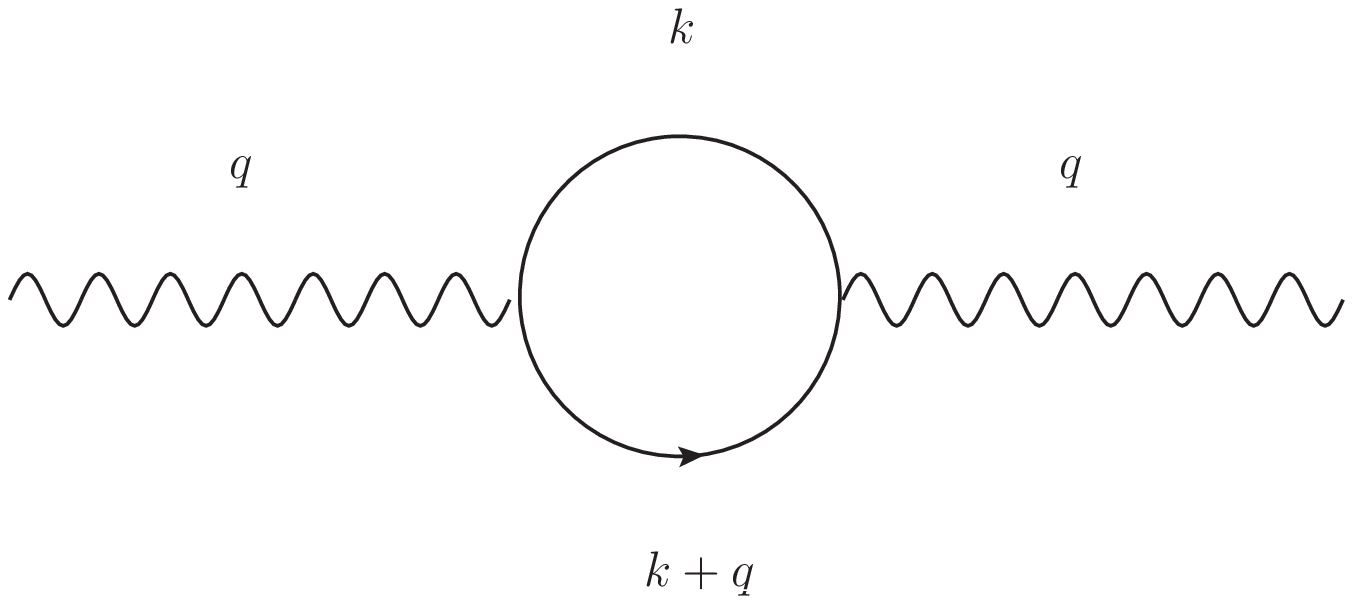}
\par\end{centering}

\begin{centering}
Fig. 1. 1-loop vacuum polarization diagram
\par\end{centering}

\end{figure}
\begin{equation}
i\Pi_{\mu\nu}(q)=-(-ig)^{2}\int\frac{d^{4}k}{(2\pi)^{4}}\hbox{Tr}\left(\gamma_{\mu}\frac{\not k+m_{a}}{k^{2}-m_{a}^{2}}\gamma_{\nu}\frac{\not\not k+\not q+m_{b}}{(k+q)^{2}-m_{b}^{2}}\right).\label{eq:pi1}\end{equation}
 $\Pi_{\mu\nu}$ is calculated with the standard technique, only the
$k_{\mu}k_{\nu}$ terms are considered with care. After performing
the trace, Wick rotating and introducing the Feynman x-parameter the
loop momentum is shifted $(k_{E\mu}+xq_{E\mu})\rightarrow l_{E\mu}$,

\begin{equation}
\Pi_{\mu\nu}=g^{2}\int_{0}^{1}dx\int\frac{d^{4}l_{E}}{(2\pi)^{4}}\frac{2l_{E\mu}l_{E\nu}-g_{\mu\nu}\left(l_{E}^{2}+\Delta\right)-2x(1-x)q_{E\mu}q_{E\nu}+2x(1-x)g_{\mu\nu}q_{E}^{2}}{\left(l_{E}^{2}+\Delta\right)^{2}},\label{eq:pi2}\end{equation}
where $\Delta=x(1-x)q_{E}^{2}+(1-x)m_{a}^{2}+xm_{b}^{2}$. In QED
$m_{a}=m_{b}=m$ and $g=e$ it simplifies to $\Delta_{1}=x(1-x)q_{E}^{2}+m^{2}$.
Having a symmetric denominator and symmetric volume of integration
the terms linear in $l_{E\mu}$ are dropped. After changing the order
of momentum- and x-integration the loop momentum is shifted with x-dependent
values, $xq_{E\mu}$ and sum up the results during the integration.
Different shifts sums up to a meaningful result only if the shift
does not modify the value of the momentum integral (it will be discussed
in the next section). 

In QED the Ward identity tells us, that

\begin{equation}
q^{\mu}\Pi_{\mu\nu}(q)=0.\label{eq:ward}\end{equation}
The terms proportional to $q_{E}$ fulfill the Ward-identity \eqref{eq:ward}
and what remains is the condition of gauge invariance\begin{equation}
\int_{0}^{1}dx\int\frac{d^{4}l_{E}}{(2\pi)^{4}}\frac{l_{E\mu}l_{E\nu}}{\left(l_{E}^{2}+\Delta_{1}\right)^{2}}=\frac{1}{2}g_{\mu\nu}\int_{0}^{1}dx\int\frac{d^{4}l_{E}}{(2\pi)^{4}}\frac{1}{\left(l_{E}^{2}+\Delta_{1}\right)}.\label{eq:condgauge}\end{equation}
This condition appeared already in \cite{wu1,nemes}. Any gauge invariant
regulator should fulfill \eqref{eq:condgauge}. It holds in dimensional
regularization and in the momentum cutoff based on DREG of Section
2. In \cite{gu,wu1} a similar relation defined the finite or infinite
Pauli-Villars terms to maintain gauge invariance.

So far the $x$ integrals were not performed. Expanding the denominator
in $q^{2}$, the x-integration can be easily done and we arrive at
a condition for gauge invariance at each order of $q^{2}$. At order
$q^{2n}$ we get (leaving out the factor $(2\pi)^{4}$) \begin{equation}
\int d^{4}l_{E}\frac{l_{E\mu}l_{E\nu}}{\left(l_{E}^{2}+m^{2}\right)^{n+1}}=\frac{1}{2n}g_{\mu\nu}\int d^{4}l_{E}\frac{1}{\left(l_{E}^{2}+m^{2}\right)^{n}},\ \ \ \ \ n=1,2,...\label{eq:cn}\end{equation}
 The conditions \eqref{eq:cn} are valid for arbitrary $m^{2}$ mass,
so it holds for any $m^{2}=\Delta$ in two or n-point functions with
arbitrary masses in the propagators. These conditions mean that in
any gauge invariant regularization the two sides of \eqref{eq:cn}
should give the same result. We will use this condition to define
the lhs of \eqref{eq:cn} in the new improved cutoff regularization.

\section{Consistency conditions - momentum routing}

Evaluating any loops in QFT one encounters the problem of momentum
routing. The choice of the internal momenta should not affect the
result of the loop calculation. The simplest example is the 2-point
function. In \eqref{eq:pi1} there is a loop-momentum $k$, and $q$
the external momentum (see Fig. 1.) is put on one line $(k+q,k)$,
but any partition of the external momentum $(k+q+p,\ \ k+p)$ must
be as good as the original. The arbitrary shift of the loop momentum
should not change the physics. This independence of the choice of
the internal momentum gives a conditions. We will impose it on a very
simple loop integral \begin{equation}
\int d^{4}k\frac{k_{\mu}}{k^{2}-m^{2}}-\int d^{4}k\frac{k_{\mu}+p_{\mu}}{\left(k+p\right)^{2}-m^{2}}=0\label{eq:shift1}\end{equation}
which turns up during the calculation of the 2-point function. Expanding
\eqref{eq:shift1} in powers of $p$ we get a series of condition,
meaningful at $p,\ p^{3},\ p^{5}\ ...$. At linear order we arrive
at \begin{equation}
\int d^{4}k\left(\frac{p_{\mu}}{k^{2}-m^{2}}-2\frac{k_{\mu}k\cdot p}{k^{2}-m^{2}}\right)=0,\label{eq:c1}\end{equation}
which is equivalent to \eqref{eq:cn} for $n=1$. At order $p^{3}$
a linear combination of two conditions should vanish \begin{equation}
p_{\rho}p_{\alpha}p_{\beta}\int d^{4}k\left[\left(\frac{4k_{\alpha}k_{\beta}}{\left(k^{2}-m^{2}\right)^{3}}-\frac{g_{\alpha\beta}}{\left(k^{2}-m^{2}\right)^{2}}\right)g_{\mu\rho}-4k_{\mu}\left(\frac{2k_{\alpha}k_{\beta}k_{\rho}}{\left(k^{2}-m^{2}\right)^{4}}-\frac{g_{\alpha\beta}k_{\rho}}{\left(k^{2}-m^{2}\right)^{3}}\right)\right]=0.\label{eq:c3}\end{equation}
These two conditions get separated if the freedom of the shift of
the loop-momentum is considered in $\int d^{4}k\frac{k_{\mu}}{\left(k^{2}-m^{2}\right)^{2}}$.
At leading order it provides \begin{equation}
p_{\nu}\int d^{4}k\left(\frac{g_{\mu\nu}}{\left(k^{2}-m^{2}\right)^{2}}-4\frac{k_{\mu}k_{\nu}}{\left(k^{2}-m^{2}\right)^{3}}\right)=0,\label{eq:cp}\end{equation}
equivalent with \eqref{eq:cn} for $n=2$. Using \eqref{eq:cp} twice
the second part of the condition \eqref{eq:c3} connects 4 loop-momenta
nominators to 2 $k$'s. Symmetrizing the indices we get\begin{equation}
\int d^{4}k\frac{k_{\alpha}k_{\beta}k_{\mu}k_{\rho}}{\left(k^{2}-m^{2}\right)^{4}}=\frac{1}{24}\int d^{4}k\frac{g_{\alpha\beta}g_{\mu\rho}+g_{\alpha\mu}g_{\beta\rho}+g_{\alpha\rho}g_{\beta\mu}}{\left(k^{2}-m^{2}\right)^{2}}.\label{eq:c4}\end{equation}
Invariance of momentum routing provides conditions for symmetry preserving
regularization and these conditions are equivalent with the conditions
coming from gauge invariance.

\section{Gauge invariance and loop-momentum shift}

We show at one loop level that gauge invariance of the vacuum polarization
function is equivalent to invariance of a special loop integrand against
shifting the loop momentum \eqref{eq:shift1}. Consider $\Pi_{\mu\nu}$
defined in \eqref{eq:pi1}, performing the trace we get\begin{equation}
i\Pi_{\mu\nu}(q)=-g{}^{2}\int\frac{d^{4}k}{(2\pi)^{4}}\frac{k_{\mu}\left(k_{\nu}+q_{\nu}\right)+k_{\nu}\left(k_{\mu}+q_{\mu}\right)-g_{\mu\nu}\left(k^{2}+k\cdot q-m_{a}m_{b}\right)}{\left(k^{2}-m_{a}^{2}\right)\left((k+q)^{2}-m_{b}^{2}\right)}.\label{eq:pi3}\end{equation}
Specially in QED $m_{a}=m_{b}=m$ and gauge invariance requires \eqref{eq:ward},
which simplifies to\begin{equation}
iq^{\nu}\Pi_{\mu\nu}(q)=-g^{2}\int\frac{d^{4}k}{(2\pi)^{4}}\left(\frac{k_{\mu}+q_{\mu}}{\left((k+q)^{2}-m^{2}\right)}-\frac{k_{\mu}}{\left(k^{2}-m^{2}\right)}\right)=0.\label{eq:pi4}\end{equation}
This example shows that the Ward identity is fulfilled only if the
shift of the loop momentum does not change the value of the integral,
like in \eqref{eq:shift1}.

In \cite{nemes2} based on the general diagrammatic proof of gauge
invariance it is shown that the Ward identity is fulfilled if the
difference of a general n-point loop and its shifted version vanishes
\begin{equation}
-i\int d^{4}p_{1}Tr\left[\frac{i}{\not p_{n}-m}\gamma^{\mu_{n}}...\frac{i}{\not p_{1}-m}\gamma^{\mu_{1}}-\frac{i}{\not p_{n}+\not q-m}\gamma^{\mu_{n}}...\frac{i}{\not p_{1}+\not q-m}\gamma^{\mu_{1}}\right]=0.\label{eq:ward-shift}\end{equation}
We interpret \eqref{eq:pi4} and \eqref{eq:ward-shift} as a necessary
condition for gauge invariant regularizations.

\section{Consistency conditions - vanishing surface terms}

All the previous conditions are related to the volume integral of
a total derivative\begin{equation}
\int d^{4}k\frac{\partial}{\partial k^{\nu}}\left(\frac{k_{\mu}}{\left(k^{2}+m^{2}\right)^{n}}\right)=\int d^{4}k\left(\frac{k_{\mu}k_{\nu}}{\left(k^{2}+m^{2}\right)^{n+1}}-\frac{1}{2n}g_{\mu\nu}\frac{1}{\left(k^{2}+m^{2}\right)^{n}}\right),\ \ \ \ \ n=1,2,...\label{eq:surface}\end{equation}
The total derivative on the lhs leads to surface terms, which vanish
for finite valued integrals and should vanish for symmetry preserving
regularization. In our improved regularization this will follow from
new definitions. The left hand side is in connection with an infinitesimal
shift of the loop momentum $k$, it should be zero if the integral
of the term in the delimiter is invariant against the shift of the
loop momentum. The vanishing of this surface terms reproduces on the
rhs the previous conditions \eqref{eq:c1} and \eqref{eq:cn}. In
\eqref{eq:surface} starting with any odd number of $k$'s in the
nominator we end up with some conditions, three $k$'s for $n=3$
provide \eqref{eq:c4} after some algebra. Starting with even number
of $k_{\mu}$'s in the nominator on the lhs in \eqref{eq:surface}
we get relations between odd number of $k_{\mu}$'s in the nominators,
which vanish separately.

These surface terms all vanish in DREG and give the basis of DREG
respecting Lorentz and gauge symmetries. Vanishing of the surface
term is inherited to any regularization, like improved momentum cutoff,
if the identification \eqref{eq:condgauge} is understood to evaluate
integrals involving even number of free Lorentz indices, e.g. nominators
alike $k_{\mu}k_{\nu}$. Vanishing of integrals with odd number of
$k$'s in the nominator is also required by the symmetry of the integration
volume.

\section{Improved momentum cutoff regularization}

We propose a new symmetry preserving regularization based on 4-dimensional
momentum cutoff. During this improved momentum cutoff regularization
method a simple sharp momentum cutoff is introduced to calculate the
divergent scalar integrals in the end. The evaluation of loop-integrals
starts with the usual Wick rotation, Feynman parameterization and
loop-momentum shift. The only crucial modification is that the potentially
symmetry violating loop integrals containing explicitely the loop
momenta with free Lorentz indices are calculated with the identification

\begin{equation}
\frac{l_{E\mu}l_{E\nu}}{\left(l_{E}^{2}+\Delta\right)^{n+1}}\rightarrow\frac{1}{2n}g_{\mu\nu}\frac{1}{\left(l_{E}^{2}+\Delta\right)^{n}},\label{eq:rule1}\end{equation}
under the loop integrals or with more momenta using the condition
\eqref{eq:c4} or generalizations of it, like \begin{equation}
\frac{l_{E\mu}l_{E\nu}l_{E\rho}l_{E\sigma}}{\left(l_{E}^{2}+\Delta\right)^{n+1}}\rightarrow\frac{1}{4n(n-1)}\cdot\frac{g_{\mu\nu}g_{\rho\sigma}+g_{\mu\rho}g_{\nu\sigma}+g_{\mu\sigma}g_{\nu\rho}}{\left(l_{E}^{2}+\Delta\right)^{n-1}}.\label{eq:rule2}\end{equation}
 Integrals with odd number of the loop-momenta vanish identically.
These identifications guarantee gauge invariance and freedom of shift
in the loop momentum. Under any regularized momentum integrals the
identifications \eqref{eq:rule1} or generalizations like \eqref{eq:rule2}
are understood as a part of the regularization procedure for $n=1,2,..$. 

What is the relation with the standard (textbook) $k_{\mu}k_{\nu}\rightarrow\frac{1}{4}g_{\mu\nu}k^{2}$
substitution? We have to modify it in case of divergent integrals
to respect gauge symmetry, i.e to fulfill \eqref{eq:cn}. Lorentz
invariance dictates that in \eqref{eq:cn} the lhs must be proportional
to the only available tensor $g_{\mu\nu}$, i.e. \begin{equation}
l_{E\mu}l_{E\nu}\rightarrow\frac{1}{d}g_{\mu\nu}l_{E}^{2}\label{eq:perd1}\end{equation}
 can be used, where $d$ is some number to determine%
\footnote{The usual method is to calculate the trace (and get d=4), but the
trace is not well defined for divergent integrals.%
}. Now both sides of equation \eqref{eq:cn} can be calculated with
simple 4-dimensional momentum cutoff. The different powers of $\Lambda$
can be matched on the two sides, and for $n=1$ we get the following
conditions (from gauge invariance) for the value of $d$,\begin{eqnarray}
\frac{1}{d}\Lambda^{2} & \rightarrow & \frac{1}{2}\Lambda^{2},\label{eq:quad2}\\
\frac{1}{d}\ln\left(\frac{\Lambda^{2}+m^{2}}{m^{2}}\right) & \rightarrow & \frac{1}{4}\left(\ln\left(\frac{\Lambda^{2}+m^{2}}{m^{2}}\right)+\frac{1}{2}\right),\label{eq:log2}\\
\frac{1}{d} & \rightarrow & \frac{1}{4}\ \ \mathrm{for\ finite\ terms}.\label{eq:fin2}\end{eqnarray}

We see that for finite valued integrals when the Wick-rotation is
legal, the condition \eqref{eq:cn} and the rule \eqref{eq:rule1}
gives the usual $k_{\mu}k_{\nu}\rightarrow\frac{1}{4}g_{\mu\nu}k^{2}$
substitution, but for divergent cases we get back the identification
partially found by \cite{hagiwara,harada,varin} and others. Quadratic
divergence goes with $d=2$, logarithmic divergence goes with $d=4$
plus a finite term (a shift), it is the $+1$ in equation \eqref{eq:log}.
For more than 2 even number of indices generalizations of \eqref{eq:perd1}
should be used, for example in case of 4 indices the \begin{equation}
l_{E\mu}l_{E\nu}l_{E\rho}l_{E\sigma}\rightarrow\frac{1}{d(d+2)}\cdot\left(g_{\mu\nu}g_{\rho\sigma}+g_{\mu\rho}g_{\nu\sigma}+g_{\mu\sigma}g_{\nu\rho}\right)l_{E}^{4}.\label{eq:perd2}\end{equation}
substitution works.

Fulfilling the condition \eqref{eq:cn} via the substitution \eqref{eq:rule1}
the results of momentum cutoff based on DREG of section 2 are completely
reproduced performing the calculation in the physical dimensions $d=4$
\cite{fcmlambda,fcmew}. The next example shows that the new regularization
provides a robust framework for calculating loop integrals and respects
symmetries.

\section{Vacuum polarization function}

As an example let us calculate the vacuum polarization function of
Fig. 1. in a general gauge theory with fermion masses $m_{a},\ m_{b}$.
For sake of simplicity we consider only vector couplings. Performing
the trace in \eqref{eq:pi1} we get \eqref{eq:pi3}. Now we can introduce
a Feynman x-parameter, shift the loop-momentum and get \eqref{eq:pi2}
after dropping the linear terms. Generally we are interested in low
energy observables like the precision electroweak parameters and need
the first few terms in the power series of $\Pi_{\mu\nu}(q)$. Using
the rule \eqref{eq:rule1} for $n=1$ and expanding the denominator
in $q^{2}$ the scalar loop and x-integrals can be easily calculated
with a 4-dimensional momentum cutoff ($\Lambda$). The result in this
construction is automatically transverse \begin{equation}
\Pi_{\mu\nu}(q)=\frac{g^{2}}{4\pi^{2}}\left(q^{2}g_{\mu\nu}-q_{\mu}q_{\nu}\right)\left[\Pi(0)+q^{2}\Pi'(0)+...\right].\label{eq:pimunu}\end{equation}
The terms independent of the cutoff completely agree with the results
of DREG \cite{fcmew} even the logarithmic singularity can be matched
with the $1/\epsilon$ terms using \eqref{eq:log} \begin{eqnarray}
\Pi(0) & \!\!=\!\! & \frac{1}{4}(m_{a}^{2}+m_{b}^{2})-\frac{1}{2}\left(m_{a}-m_{b}\right)^{2}\ln\left(\frac{\Lambda^{2}}{m_{a}m_{b}}\right)-\label{eq:pi0}\\
 &  & \!\!-\!\!\frac{m_{a}^{4}+m_{b}^{4}-2m_{a}m_{b}\left(m_{a}^{2}+m_{b}^{2}\right)}{4\left(m_{a}^{2}-m_{b}^{2}\right)}\ln\left(\frac{m_{b}^{2}}{m_{a}^{2}}\right).\nonumber \end{eqnarray}
 The first derivative is \begin{eqnarray}
\Pi'(0) & \!\!=\!\!\! & -\frac{2}{9}-\frac{4m_{a}^{2}m_{b}^{2}-3m_{a}m_{b}\left(m_{a}^{2}+m_{b}^{2}\right)}{6\left(m_{a}^{2}-m_{b}^{2}\right)^{2}}+\frac{1}{3}\ln\left(\frac{\Lambda^{2}}{m_{a}m_{b}}\right)+\label{eq:piv0}\\
 &  & +\frac{\left(m_{a}^{2}+m_{b}^{2}\right)\left(m_{a}^{4}-4m_{a}^{2}m_{b}^{2}+m_{b}^{4}\right)+6m_{a}^{3}m_{b}^{3}}{6\left(m_{a}^{2}-m_{b}^{2}\right)^{3}}\ln\left(\frac{m_{b}^{2}}{m_{a}^{2}}\right).\nonumber \end{eqnarray}
The photon remains massless in QED, as in the limit, $m_{a}=m_{b}$
we get $\Pi(0)=0$.

In this paragraph we show that the proposed regularization is robust
and gives the same result even if the calculation is organized in
a different way. Introducing Feynman parameters and shifting the loop
momentum can be avoided if we need only the first few terms in the
Taylor expansion of $q$. For small $q$ the second denominator in
\eqref{eq:pi3} can be Taylor expanded, for simplicity we give the
expanded integrand for equal masses, up to ${O}(q^{4})$\begin{eqnarray}
\Pi_{\mu\nu}(q) & \simeq & -g{}^{2}\int\frac{d^{4}k_{E}}{(2\pi)^{4}}\left[2k_{\mu}k_{\nu}\left(\frac{1}{\left(k_{E}^{2}+m^{2}\right)^{2}}-\frac{q_{E}^{2}}{\left(k_{E}^{2}+m^{2}\right)^{3}}+\frac{4\left(k_{E}\cdot q_{E}\right)^{2}}{\left(k_{E}^{2}+m^{2}\right)^{4}}\right)\right.\label{eq:piexp}\\
 & -\negmedspace\! & \left.\frac{2\left(k_{E\mu}q_{E\nu}+k_{E\nu}q_{E\mu}\right)k_{E}\cdot q_{E}}{\left(k_{E}^{2}+m^{2}\right)^{3}}-g_{\mu\nu}\left(\frac{1}{\left(k_{E}^{2}+m^{2}\right)^{2}}-\frac{q_{E}^{2}}{\left(k_{E}^{2}+m^{2}\right)^{3}}+\frac{2\left(k_{E}\cdot q_{E}\right)^{2}}{\left(k_{E}^{2}+m^{2}\right)^{4}}\right)\right].\nonumber \end{eqnarray}
 Taking into account that $k_{E}\cdot q_{E}=k_{E\alpha}q_{E\alpha}$,
\eqref{eq:rule1} and \eqref{eq:rule2} can be used and the remaining
scalar integrals can be easily calculated. The result agrees with
\eqref{eq:pi0} and \eqref{eq:piv0} and the finite terms with DREG
if and only if we use the correct symmetry preserving substitutions.
Applying the naive $\frac{1}{4}g_{\mu\nu}k_{E}^{2}$ substitution
in both approaches the finite terms will differ not just from each
other but also from the result of DREG.

The calculation of the $\Pi_{\mu\nu}$ function at 1-loop shows that
the new regularization gives a robust gauge invariant result and the
finite terms agree with DREG.

\section{Conclusions}

We presented in this paper a new method for the reliable calculation
of divergent 1-loop diagrams with four dimensional momentum cutoff.
Various conditions were derived to maintain gauge symmetry, to have
the freedom of momentum routing or shifting the loop-momentum. These
conditions were known by several authors \cite{gu,wu1,nemes,nemes2}.
Our new proposal is that these conditions will be satisfied during
the regularization process if terms proportional to loop-momenta with
free Lorentz indices (e.g. $\sim k_{\mu}k_{\nu}$) are calculated
according to the special rules \eqref{eq:rule1} and \eqref{eq:rule2}
or generalizations thereof. In the end the scalar integrals are calculated
with a simple momentum cutoff. The calculation is robust - at least
at 1-loop level - as we have shown via the fermionic contribution
to the vacuum polarization function. The finite terms agree with the
one in DREG in all examples. The connection with DREG is more transparent
if one uses alternatively the $k_{\mu}k_{\nu}\rightarrow\frac{1}{d}g_{\mu\nu}k^{2}$
or \eqref{eq:perd2} substitution and $d$ has different values determined
by the degree of divergence in each term (\ref{eq:quad2}, \ref{eq:log2},
\ref{eq:fin2}). $d=2$ for quadratic divergencies, for log divergent
terms $d=4$ further there is an important finite shift, and simply
$d=4$ for finite terms. The new improved momentum cutoff regularization
at 1-loop gives the same results as the cutoff regularization based
on DREG of section 2. This observation gives a solid basis to use
the new method for complicated diagrams or at higher loops, finite
terms are expected to agree with DREG. We stress that this new regularization
stands without DREG as the substitutions \eqref{eq:rule1}, \eqref{eq:rule2}
and scalar integration with a cutoff are independent of DREG. The
success of both regularizations based on the property that they fulfill
the consistency conditions of gauge invariance and momentum shifting.

The idea to use consistency conditions has been tested in the literature
by various authors, we list few examples. The infinite terms in a
cutoff calculation using \eqref{eq: perdim} were identified correctly
in \cite{varin}, the authors showed that the 1-loop QED Ward identities
are fulfilled and the Goldstone theorem is recovered in the phenomenological
chiral model. Constrained differential renormalization proved to be
useful also in supersymmetric \cite{drsusy} and non-Abelian gauge
theories, it fulfills Slavnov-Taylor identities at one and two loops
\cite{drym}. Implicit regularization \cite{nemes,nemes2} requires
the same conditions as we used and it was successfully applied to
the Nambu-Jona-Lasinio model \cite{nemes} and to higher loop calculations
in gauge theory. It was shown that the conditions guarantee gauge
invariance generally and the Ward identities are fulfilled explicitely
in QED at two-loop order \cite{nemes2}. In an effective composite
Higgs model, the Fermion Condensate Model \cite{fcm} oblique radiative
corrections (S and T parameters) were calculated in DREG and with
the improved cutoff, the finite results completely agree. The calculation
involved vacuum polarization functions with two different fermion
masses and no ambiguity appeared \cite{fcmlambda,fcmew}. 

This new regularization prescription is advantageous in special loop-calculations
where one wants to keep the cutoff of the model, like in effective
theories, derivation of renormalization group equations, extra dimensional
scenarios or in models explicitely depending on the space-time dimensions
like supersymmetric theories. We argue that the method can be successfully
used in higher order calculations containing terms up to quadratic
divergencies in (non-Abelian) gauge theories as it allows for shifts
in the loop momenta, which guarantees the 't Hooft identity \cite{nemes2,thooft}.
This symmetry preserving method can be used also in automatized calculations
(similar to \cite{autom}) as even the Veltman-Passarino functions
\cite{vp} can be defined with the improved cutoff.

\appendix

\section{Basic integrals}

In this appendix we list the basic divergent integrals calculated
by the regularization proposed in this paper. In the following formulae
$m^{2}$ can be any loop momentum $(k)$ independent expression depending
on Feynman x parameter, external momenta, etc., e.g. $\Delta(x,q,m_{a},m_{b}).$
\begin{eqnarray}
\int_{0}^{\Lambda}\frac{d^{4}k}{i(2\pi)^{4}}\frac{1}{k^{2}-m^{2}} & \!\!=\!\! & -\frac{1}{(4\pi)^{2}}\left(\Lambda^{2}-m^{2}\ln\left(\frac{\Lambda^{2}}{m^{2}}\right)\right)\label{eq:i1}\\
\int_{0}^{\Lambda}\frac{d^{4}k}{i(2\pi)^{4}}\frac{k_{\mu}k_{\nu}}{\left(k^{2}-m^{2}\right)^{2}} & \!\!=\!\! & -\frac{1}{(4\pi)^{2}}\frac{g_{\mu\nu}}{2}\left(\Lambda^{2}-m^{2}\ln\left(\frac{\Lambda^{2}}{m^{2}}\right)\right)\label{eq:i2}\\
\int_{0}^{\Lambda}\frac{d^{4}k}{i(2\pi)^{4}}\frac{k_{\mu}k_{\nu}k_{\rho}k_{\sigma}}{\left(k^{2}-m^{2}\right)^{3}} & \!\!=\!\! & \!\!-\frac{1}{(4\pi)^{2}}\frac{g_{\mu\nu}g_{\rho\sigma}+g_{\mu\rho}g_{\nu\sigma}+g_{\mu\sigma}g_{\nu\rho}}{8}\left(\Lambda^{2}-m^{2}\ln\left(\frac{\Lambda^{2}}{m^{2}}\right)\right)\label{eq:i3}\\
\int_{0}^{\Lambda}\frac{d^{4}k}{i(2\pi)^{4}}\frac{k^{2}k_{\mu}k_{\nu}}{\left(k^{2}-m^{2}\right)^{3}} & \!\!=\!\! & -\frac{1}{(4\pi)^{2}}\frac{g_{\mu\nu}}{4}\left(\Lambda^{2}-2m^{2}\ln\left(\frac{\Lambda^{2}}{m^{2}}\right)+\frac{1}{2}m^{2}\right)\label{eq:i3a}\\
\int_{0}^{\Lambda}\frac{d^{4}k}{i(2\pi)^{4}}\frac{1}{\left(k^{2}-m^{2}\right)^{2}} & \!\!=\!\! & \frac{1}{(4\pi)^{2}}\left(\ln\left(\frac{\Lambda^{2}}{m^{2}}\right)-1\right)\label{eq:i4}\\
\int_{0}^{\Lambda}\frac{d^{4}k}{i(2\pi)^{4}}\frac{k_{\mu}k_{\nu}}{\left(k^{2}-m^{2}\right)^{3}} & \!\!=\!\! & \frac{1}{(4\pi)^{2}}\frac{g_{\mu\nu}}{4}\left(\ln\left(\frac{\Lambda^{2}}{m^{2}}\right)-1\right)\label{eq:i5}\\
\int_{0}^{\Lambda}\frac{d^{4}k}{i(2\pi)^{4}}\frac{k^{2}k_{\mu}k_{\nu}}{\left(k^{2}-m^{2}\right)^{4}} & \!\!=\!\! & \frac{1}{(4\pi)^{2}}\frac{g_{\mu\nu}}{6}\left(\ln\left(\frac{\Lambda^{2}}{m^{2}}\right)-\frac{3}{2}\right)\label{eq:i5a}\\
\int_{0}^{\Lambda}\frac{d^{4}k}{i(2\pi)^{4}}\frac{k_{\mu}k_{\nu}k_{\rho}k_{\sigma}}{\left(k^{2}-m^{2}\right)^{4}} & \!\!=\!\! & \frac{1}{(4\pi)^{2}}\frac{g_{\mu\nu}g_{\rho\sigma}+g_{\mu\rho}g_{\nu\sigma}+g_{\mu\sigma}g_{\nu\rho}}{24}\left(\ln\left(\frac{\Lambda^{2}}{m^{2}}\right)-1\right)\label{eq:i6}\end{eqnarray}
(\ref{eq:i1}-\ref{eq:i3}) depend on the same function of $\Lambda$
as (\ref{eq:i2}, \ref{eq:i3}) are traced back to \eqref{eq:i1}
via \eqref{eq:rule1} and \eqref{eq:rule2}. On the other hand \eqref{eq:i3a}
has a different $\Lambda$ dependence showing that \eqref{eq:rule1}
or \eqref{eq:rule2} applies only to free indices, contraction of
Lorentz indices does not commute with the integration in case of divergencies.

\end{document}